\documentclass[a4paper,12pt]{article}
\usepackage{verbatim}        
\usepackage{graphicx,subfigure,epsfig,float}
\usepackage{xspace}
\usepackage{xcolor}
\usepackage{array,multirow}
\usepackage{amsmath,amssymb}
\usepackage{xspace}
\usepackage{amsmath,amssymb,amsfonts}

\newcommand{\bs}[1]{\ensuremath{\boldsymbol{#1}}}

\newcommand{\bo}{\ensuremath{\bs o}\xspace}
\newcommand{\bp}{\ensuremath{\bs p}\xspace}

\newcommand{\alp}{\ensuremath{\alpha}\xspace}

\renewcommand{\Re}{\ensuremath{\mathbb{R}}\xspace} 

\newcommand{\norm}[1]{\ensuremath{\| #1\|}\xspace}

\newcommand{\refeq}[1]{Eq.~(\ref{eq:#1})\xspace}

\begin{document}
%
%
\title{Introducing Barik transform and Barik eliminant: novel tools for solving polynomial equations in the sine and cosine of one variable}
\author{Sandipan Bandyopadhyay\thanks{Email: 
		sandipan@iitm.ac.in.}\\
Department of Engineering Design\\
Indian Institute of Technology Madras}
\date{}
\maketitle
\begin{abstract}
	Systems of trigonometric equations appear frequently in problems related to geometry, kinematics, computer graphics, robotics, and astronomy, among many other fields. Most often these are first converted to algebraic equations via the ``tangent half-angle'' substitution and solved subsequently using various algebraic methods, such as computation of resultants or Gr\"obner bases. In the process, parametric singularities are introduced in the equations, and their degrees are also doubled. In this article, a novel concept called the ``Barik transform'' is introduced, which can help solve such a system without encountering these issues, albeit only in systems that contain {\em at least} one equation of the first degree in the sine and cosine of the variable to be eliminated. It is shown how Barik transform can produce extremely compact eliminants in the closed form, without resorting to any resultant computation. 	
\end{abstract}
\section{Introduction}
\label{sc:intro}
The  concept of {\em Barik} (pronounced ``Baa-reek'')  transform\footnote{The transform is so named to honour the memory of the late Professor Panchanan Barik, Professor of Physics at Serampore College, West Bengal, India, who taught the author during his high-school days.} was recently introduced by the author, in the context of solution of ``trigonomial''\footnote{The term ``trigonomial'' is used to abbreviate the phrase ``polynomial equation of trigonometric functions (such as the sine and cosine of any variable)''.} equations. Although it has already been used in a few recent studies, e.g.,~\cite{kolte2024, patra2024, patra2025, firat2025,shashank2026, patel2026}, a detailed discussion on it has not been published yet. 

Barik transform is a mathematical tool to convert trigonomial equations into their algebraic counterparts. A very well-established method for the same is the so-called ``tangent half-angle''~transformation (abbreviated as ``HTT'' hereinafter for the sake of brevity), which uses the following identities (see, e.g.,~\cite{roth96}):
\begin{align}
	\sin\theta = \frac{2t}{1+t^2},  \quad
	\cos\theta = \frac{1-t^2}{1+t^2}, \quad 
	\text{where} \ t = \tan(\theta/2).
\end{align}
It is obvious that this transformation introduces a {\em parametric singularity} into the subsequent analysis at~$\theta = \pm \pi$, since~$t$ approaches~$\pm \infty$, simultaneously, at these points. This undesirable artefact of using this transformation had been pointed out decades back~\cite{kovacshommel1993}. However, the convenience it affords has allowed it to not just survive, but rather thrive in the domains of kinematics and geometry, giving it the status of the {\em de facto} standard method for such problems. Another consequence of the HTT is that it raises the degree of the resulting polynomial(s). In particular, if a given trigonomial (in the sine and cosine of a single variable) has a total degree of~$n$, then the degree of the corresponding polynomial equation in~$t$ (defined as above) will, in general, be~$2n$. Thus, the advantages of converting the original transcendental equation to an algebraic one is derived only at the cost of doubling the degree of the resulting equation. 

Given these undesirable consequences of the~HTT, it seems logical to develop alternatives without these drawbacks. While it may not be possible to alleviate these issues {\em globally}, it is possible to circumvent them at least in certain situations. Barik transform presents such a remedy, albeit in a situation involving a system of trigonomials in the same variable, wherein {\em at least} one of the equations is linear in the sine and cosine of the said variable. 
\section{Derivation of Barik transform}
\label{sc:bariktrans}
\begin{figure}[t]
	\centering
	\includegraphics[width = 0.6\textwidth] {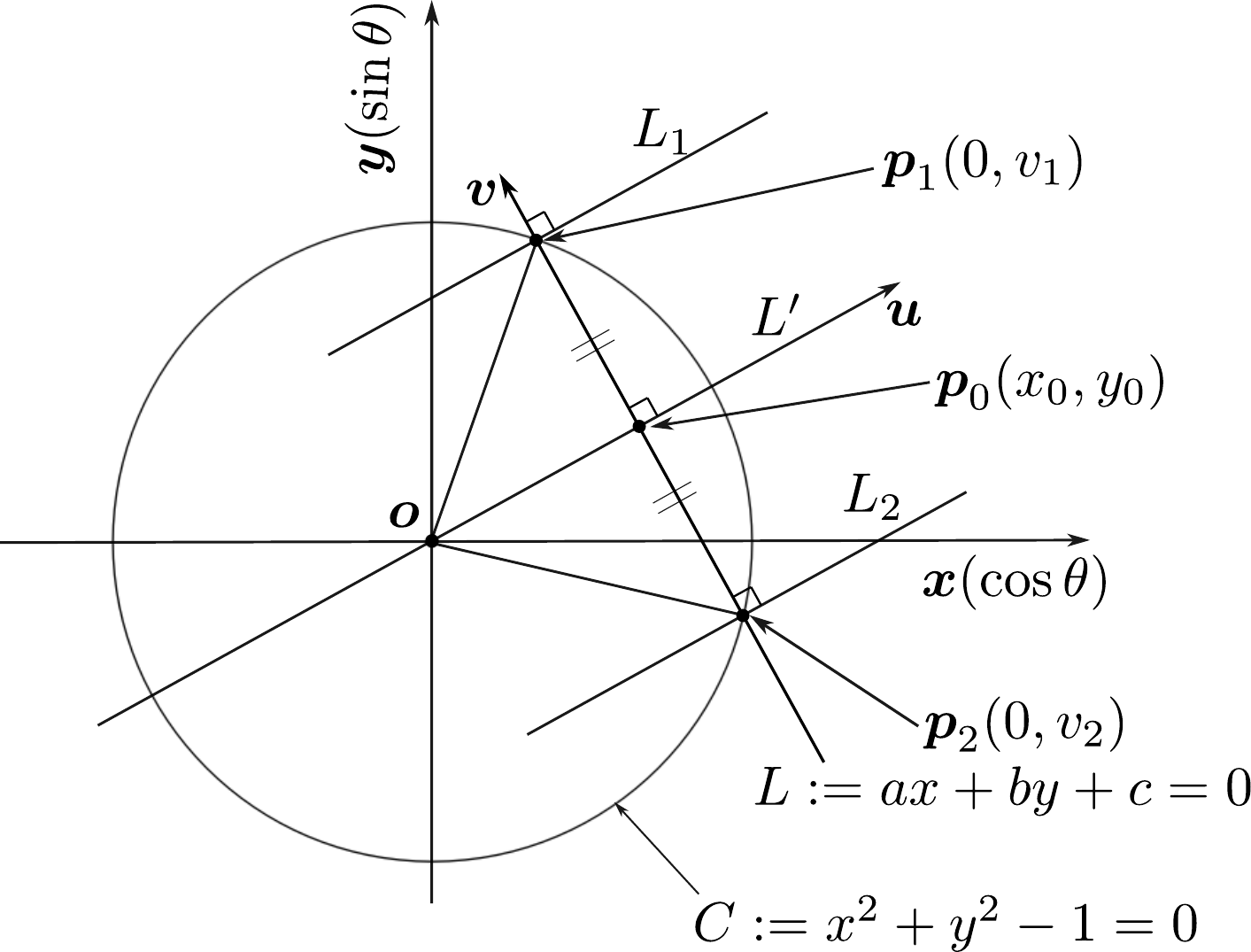}
	\caption{Geometric interpretations of \refeq{lintrig}}
	\label{fg:cirline}
\end{figure}
The concept of Barik transform (abbreviated as ``BT'' hereinafter) is explained below in detail. 
As mentioned in Section~\ref{sc:intro}, BT applies to a system of trigonomials only when at least one of the equations of the system is of degree one in the sine and cosine of a trigonometric variable. In its generic form, such an equation can be represented as:
\begin{align}
	\label{eq:lintrig}
	& f(\theta) = a \cos\theta + b\sin \theta + c, \quad 
	a, b, c \in \mathbb{R}. 
\end{align}
Equation~(\ref{eq:lintrig}) represents a line, say,~$L$, in the 2-D space spanned by~$\{\sin\theta, \cos\theta\}$ (see Fig.~\ref{fg:cirline}). However, the number of {\em real} points on this line, {\em at the most}, can only be {\em two}. These  points, when they exist, lie at the intersection of the line~$L$ with the unit circle~$C$ centred at the origin, the equation of which is given by the ancient identity: 
\begin{align}
	\cos^2\theta + \sin^2 \theta = 1.
	\label{eq:circle}
\end{align}
For the sake of brevity, let~$x = \cos\theta$ and~$y = \sin\theta$. Therefore, Eq.~(\ref{eq:lintrig}) may be rewritten as: 
\begin{align}
	\label{eq:line2}
	a x + b y + c = 0.
\end{align}
Similarly, \refeq{circle} becomes: 
\begin{align}
	\label{eq:circle2}
	x^2 + y^2 = 1.
\end{align}
As mentioned before, the task is to find the points of intersection of the circle~$C$ and the line~$L$. When the solutions are real and distinct, the line~$L$ carries a chord\footnote{If the roots coincide then~$L$ becomes a tangent to the circle.} of the circle~$C$. It is obvious that the triangle~$\bo\bp_1\bp_2$ is isosceles in nature. To explore this two-way symmetry further, the perpendicular bisector of the chord, namely,~$L'$, is considered. The line~$L'$ passes through the origin~\bo, and intersects~$L$ at the point~$\bp_0$ (say). In doing so, the line~$L'$ divides the  triangle~$\bo\bp_1\bp_2$ into two equal halves, which are reflections of each other about~$L'$. This symmetry plays the most crucial role in the derivations that follow. 

The line~$L'$ has the equation: 
\begin{align}
	\label{eq:L4}
	& bx - ay = 0.
\end{align}	
The coordinates of~$\bp_0$ may be found as:
\begin{align}
	\label{eq:pstar}
	\bp_0 = [x_0, y_0]^\top =  \left(\frac{-c}{a^2+b^2}\right)[a,b]^\top, 
	\quad \text{assuming}\ a^2 + b^2 \ne 0. 
\end{align}
The following notations are introduced at this point:
\begin{align}
	& \rho = a^2 + b^2, \quad \text{and} \nonumber \\
	& w = -\frac{c}{a^2 + b^2} = -\frac{c}{\rho}.
	\label{eq:defw}
	\intertext{With these, \refeq{pstar} reduces to:}
	& \bp_0 = w[a,b]^\top.
\end{align}
For the points of intersection,~$\bp_1,\bp_2$ to be real, the point~$\bp_0$ must lie {\em inside} the circle~$C$, or  in the limiting case, {\em on} it; i.e.:
\begin{align}
	& \norm{\bp_0}^2 = w^2(a^2+b^2)  = w^2 \rho\le 1. \nonumber \\
	\Rightarrow & \frac{c^2}{\rho} \le 1 \nonumber,
	\quad \text{since} \ w = -\frac{c}{\rho}; 
	\nonumber\\
	\Rightarrow	& \Delta \ge 0, 
	\quad \text{where} 
	\ \Delta = \rho - c^2  = a^2 + b^2 - c^2. 
	\label{eq:defdelta}
\end{align}

A new frame of reference is constructed next, in which~$\bp_0$ serves as the origin, and the new coordinates,~$\{u,v\}$, locate a point along~$L'$ and~$L$, respectively. This is achieved by the following linear transformation: 
\begin{align}
	\label{eq:transform}
	\begin{bmatrix}
		x \\ y
	\end{bmatrix}	
	& = 
	\begin{bmatrix}
		x_0 \\ y_0
	\end{bmatrix}	
	+
	\begin{bmatrix}
		a & -b \\
		b & a	
	\end{bmatrix}	
	\begin{bmatrix}
		u \\ v
	\end{bmatrix} \nonumber \\
		& = w
	\begin{bmatrix}
		a \\ b
	\end{bmatrix}	
	+
	\begin{bmatrix}
		a & -b \\
		b & a	
	\end{bmatrix}	
	\begin{bmatrix}
		u \\ v
	\end{bmatrix}.
\end{align}
It may be noted that the map between~$\{x,y\}$ and~$\{u,v\}$ is one-to-one and onto, since as per the usual assumptions\footnote{It is assumed that~$a,b$ cannot vanish simultaneously, since in that case \refeq{lintrig} becomes either inconsistent (when~$c \ne 0$) or reduces to an identity (if~$c = 0$).}, the determinant of the transformation matrix on the RHS of \refeq{transform} equals~$a^2 + b^2>0$. 
Further, the affine transformation presented in \refeq{transform} may be recognised as an instance of the more generic family of {\em Helmert} transformations, wherein~$a^2 + b^2$ does not necessarily equal unity\footnote{When~$a^2 + b^2 = 1$ \refeq{transform} represents a rigid motion in the plane.}. 

Equation~(\ref{eq:transform}) can be simplified by noting that the points of interest lie on the line~$L$, which is given by~$u= 0$: 
\begin{align}
	\begin{bmatrix}
		x \\ y
	\end{bmatrix}
	& = w
	\begin{bmatrix}
		a \\ b
	\end{bmatrix}	
	+
	\begin{bmatrix}
		a & -b \\
		b & a	
	\end{bmatrix}	
	\begin{bmatrix}
		0 \\ v
	\end{bmatrix}.
\end{align}
Finally,~$[x,y]^\top$ is related to~$[w,v]^\top$ via a {\em linear} transformation:
\begin{align}
	\label{eq:barik}
\begin{bmatrix}
	x \\ y
\end{bmatrix}
&=
\begin{bmatrix}
	a & -b \\
	b & a	
\end{bmatrix}	
\begin{bmatrix}
	w \\ v
\end{bmatrix}.
\end{align}
Equation~(\ref{eq:barik}) is termed as the {\em linear form} of the Barik transform. It expresses the sine and cosine of~$\theta$ algebraically in terms of the parameters~$w, v$, of which~$w$ has already been expressed analytically in terms of the coefficients of the original trigonomial. It remains to find~$v$ in a similar manner. 

To determine~$v$, \refeq{barik} is used in the equation of the circle, \refeq{circle2}, expressing it in the new coordinates: 
\begin{align}
	\label{eq:newcircle}
	\left(aw - bv\right)^2 + \left( bw + av\right)^2 = 1,
\end{align}
Equation~(\ref{eq:newcircle}) looks more complicated than its original counterpart. However, it can be readily simplified to: 
\begin{align}
	\label{eq:vsqr}
	& (a^2 + b^2)(v^2 + w^2) = 1;	 \nonumber \\
	\Rightarrow & v^2 = \frac{a^2 + b^2 - c^2}{\left(a^2 + b^2\right)^2}; \nonumber \\
	\Rightarrow & v^2 = \frac{\Delta}{\rho^2}
	\quad \text{(from \refeq{defdelta})}. 
\end{align}
The solutions for~$v$ can be obtained in the closed form from \refeq{vsqr}. These can be expressed compactly as:
\begin{align}
	\label{eq:v1v2}
	v_{1,2} = \pm \frac{\sqrt{\Delta}}{\rho}.
\end{align}
Thus, the points of intersection,~$\bp_1, \bp_2$ can be expressed in the new coordinates as: 
\begin{align}
	\label{eq:p1p2}
		& \bp_i = [0, v_i]^\top, \quad i = 1,2.
\end{align}
Using Eq.~(\ref{eq:barik}) one can easily obtain the corresponding coordinates,~$\{x_i,y_i\}$, in the original frame of reference:
 \begin{align}
 	\label{eq:barik2}
 	\begin{bmatrix}
 		x_i \\ y_i
 	\end{bmatrix}
 	&=
 	\begin{bmatrix}
 		a & -b \\
 		b & a	
 	\end{bmatrix}	
 	\begin{bmatrix}
 		w \\ v_i
 	\end{bmatrix}, \ i = 1,2.
 \end{align}

Finally, the solutions for~$\theta$ are obtained as:
\begin{align}
	\label{eq:soltheta2}
	& \theta_i = \operatorname{atan2}(y_i, x_i), \quad i = 1,2,
\end{align}
where~$\operatorname{atan2}(r\sin(\cdot), r\cos(\cdot))$, $r \in \Re\backslash\{0\}$, is the {\em four-quadrant} arctangent function. 
\section{Application of Barik transform in solving systems of trigonomial equations: Barik eliminant}
\label{sc:applications}
As mentioned before, Barik transform can be used to convert trigonomial equations into their respective polynomial counterparts, thereby facilitating their solution process. This is demonstrated below in the context of a system of trigonomials in the single variable,~$\theta$, consisting of only two equations\footnote{This restriction does not diminish the generality of the formulation; see~\cite{kolte2024} for an application involving a system of four trigonomials.}:
\begin{align}
	\label{eq:f}
	& f(\theta) = a\cos\theta + b \sin\theta + c,
	\quad a,b, c \in \Re; \quad \text{and} \\
	\label{eq:g}
	& g(\theta) = 0,
\end{align}
where~$g$ has a total degree of~$m \in \mathbb{Z}^+$ (in~$\sin\theta$ and~$\cos\theta$). The coefficients of~$g$ are real constants. 
The steps are described in detail below. 
\begin{enumerate}
	\item Applying Barik transform to \refeq{f}, one obtains {\em linear} parametrisations of~$\sin\theta$ and~$\cos\theta$ as:
	\begin{align}
		\label{eq:sincos}
		& \cos\theta = aw - bv \nonumber \\
		& \sin\theta = bw + av, \quad \text{where}\
		w = -\frac{c}{a^2 + b^2} = -\frac{c}{\rho} 
		\quad \text{(see \refeq{defw})}.
	\end{align}
	Substituting these expressions in \refeq{g}, it is converted to a {\em univariate polynomial} of degree~$m$ in~$v$, which is expressed as:
	\begin{align}
		\label{eq:gtilde}
		\tilde{g}(v) = 0. 
	\end{align}
	\item Similarly, using \refeq{sincos} in the equation of the unit circle, it is transformed into a quadratic in the unknown parameter,~$v$:
	\begin{align}
		\label{eq:vtilde}
		& \tilde{f}(v) := \rho(w^2 + v^2) - 1 = 0.
	\end{align}
	As shown in \refeq{vsqr} before,~$v^2$ can be obtained explicitly as~$v^2= \frac{\Delta}{\rho^2}$.
	\item 
	The resulting polynomial {\em ideal}~$\langle\tilde{f},  \tilde{g}\rangle$ constitutes an over-constrained system in a single unknown,~$v$. Therefore, the next step is to derive the consistency condition, such that this system can possess at least one common root in~$v$. This is easily achieved by substituting~$v^2 = \frac{\Delta}{\rho^2}$ directly into~$\tilde{g}(v) = 0$, such that it is reduced to (at the most) a {\em linear} equation in~$v$, say, of the following 
	form:
	%
	\begin{align}
		\label{eq:h}
		&h(v) := \alp v + \beta = 0. 
	\intertext{Therefore, the common solution is obtained in the {\em closed form}:}
	\label{eq:defv}
	&v = -\beta/\alpha, \quad \text{assuming} 
	\ \alpha \ne 0.  
	\intertext{Finally, the desired consistency condition is obtained by squaring both sides of \refeq{defv}, and using \refeq{vsqr}:}
	\label{eq:elim}
	& \alpha^2 \Delta - \beta^2 \rho^2= 0. 
	\end{align}
	Thus, \refeq{elim} defines the {\em Barik eliminant} of~$v$ from~$\langle\tilde{f},  \tilde{g}\rangle$, and equivalently, the eliminant of~$\theta$ from the original trigonomial equations, $f(\theta) =0, g(\theta) = 0$. 
\end{enumerate}
	It is important to note a few points at this stage. 
	\begin{enumerate}
		\item The above derivation is invariant w.r.t.~$m$, i.e., the degree of the larger polynomial. Irrespective of the degree, the common solution, as well as the eliminant, can {\em always} be obtained in closed forms, via only basic algebraic operations. The degree and coefficients of~$g(\theta)$ would only impact the actual expressions of~$\alp$ and~$\beta$ in \refeq{elim}.
		\item It is not necessary to substitute the expression of~$v^2$ in \refeq{gtilde} to reduce it to a linear equation in~$v$. The same effect could have been achieved, e.g., by a polynomial division, of~$\tilde{g}(v)$ by~$\tilde{f}(v)$, w.r.t.~$v$. 
		\item In the same spirit, it is not necessary to derive~\refeq{h}, either by substitution or via polynomial division. One could eliminate~$v$ from~\refeq{vtilde} and~\refeq{gtilde} directly by computing the resultant of these polynomials w.r.t~$v$. However, that would lead to much more demanding symbolic computations, whose difficulties have been summarised in~\cite{kolte2024}. In fact, even if~$m$ is relatively small, say, $6$ or~$8$, it may become difficult, or even impossible in practice, to derive the eliminant symbolically in the closed form. In the proposed method, however, it is {\em always} possible to compute~$\alp, \beta$ in the closed form, and hence, a very compact form of the eliminant can always be achieved symbolically, in terms of the coefficients of the original trigonomials. 
		\item It may appear that the formulation would break down at~$\alpha = 0$. However, \refeq{h} shows that if~$\alpha$ vanishes, so does~$\beta$. Therefore,~$\tilde{g}$ is the GCD of~$\tilde{f}, \tilde{g}$, up to a constant factor. It may be obtained accordingly. 
	\end{enumerate}

\section{Discussions}
\label{sc:disc}
This brief article serves the purpose of introducing Barik transform and its application in solving trigonomial equations, even though these concepts have already found application in a handful of research work. It is not proposed as a global alternative to the more familiar HTT;
nevertheless, its domain of application is not very small either. It is very common in mechanisms and robotics to encounter a system of trigonomial equations among which at least one is linear in the sine and cosine of a single variable, as seen in the references cited in this article. In such situations Barik transform not just alleviates the problem of parametric singularities inherent in HTT, but also provides a means of eliminating variable(s) via significantly lesser computational burden. In particular, there is no need to derive the resultant of polynomials in their symbolic forms and subsequently computing the nullspaces of matrices numerically. It is hoped that these desirable features of Barik transform and the computational simplicity of the associated elimination process would facilitate analytical research in robotics, mechanisms and allied fields. 

As may be expected intuitively, Barik transform can be related to a number of other geometric and algebraic concepts. For instance, the direct symbolic solution for the trigonometric variable appearing in the linear equation (of sine and cosine of it) may be readily obtained from the same construction as in Fig.~\ref{fg:cirline}. On the other hand, from an algebraic perspective, it can be seen that $\tilde{g}(v) \equiv h(v) \ (\mod \tilde{f}(v))$. Hence, the construction of the algebraic eliminant may also be explained in that context. These additional details would be included in the future revisions of this document. 
\bibliographystyle{ieeetr}
\bibliography{barikref}
\end{document}